\documentstyle[here,psfig,referee]{mn}
\begin{document}
\def\simlt{\mathrel{\rlap{\lower 3pt\hbox{$\sim$}}\raise 2.0pt\hbox{$<$}}}
\def\simgt{\mathrel{\rlap{\lower 3pt\hbox{$\sim$}} \raise 2.0pt\hbox{$>$}}}
\def\di{\mbox{d}}
\def\Msun{M_{\odot}}
 
\newcommand{\q}{\begin{equation}}
\newcommand{\qa}{\begin{eqnarray}}
\newcommand{\qs}{\begin{eqnarray*}}
\newcommand{\nq}{\end{equation}}
\newcommand{\nqa}{\end{eqnarray}}
\newcommand{\nqs}{\end{eqnarray*}}

\title[IS PRIMORDIAL $^4$HE TRULY FROM BIG BANG ?]{IS PRIMORDIAL $^4$HE TRULY FROM BIG BANG ?}

\author[Ruben Salvaterra \& Andrea Ferrara]{R. Salvaterra \& A. Ferrara \\
SISSA/International School for Advanced Studies,
Via Beirut 4, 34100 Trieste, Italy}
\maketitle \vspace {3cm }
 
\begin{abstract}
Population III stars are now believed to contribute to the observed 
Near Infrared Background
(NIRB) and heavy element pollution of the intergalactic medium.
Here we show that a Pop III contribution to the
primordial $^4$He abundance consistent with NIRB constraints,        
might mask a lower value for the Big Bang Nucleosynthesis (BBN) 
$^4$He abundance.
\end{abstract}                                                                  

\smallskip
\noindent
As observations start to explore cosmic epochs close to the formation
of the first, so-called Population III (Pop III) stars, 
theoretical models provide increasingly detailed 
predictions on the properties and evolution of these objects. 
Bond, Carr \& Arnett (1983) and Carr (1994),  
discussed a possible contribution  of Pop III stars  to the 
primordial $^4$He mass fraction (Y$_p$). 
The observational data at that time did not allow to check 
the validity of their hypothesis. Current experiments are able to 
determine Y$_p$ with a precision $\simlt$ 1\% allowing  further 
investigations of the problem. 
Such experiments are based on high signal-to-noise ratio spectra 
of a class of star-forming
galaxies called Blue Compact Dwarf (BCD) galaxies. These are low-luminosity
($M_B\ge -18$) systems undergoing an intense burst of star formation occurring
in a very compact region (less than 1 kpc)  
characterized by a blue color and a HII region-like emission-line optical
spectrum, which dominates the galactic light. BCDs are among the most metal-deficient
gas-rich galaxies known. Their gas has not been processed through many
generation of stars, and thus best approximates the pristine primordial composition.
Thus, Y$_p$ can be derived accurately with only
a small correction for helium produced in stars. Moreover, the theory of nebular
emission is understood well enough to convert $^4$He emission-line
strengths into abundances with the desired accuracy.
Y$_p$ is generally determined \cite{Peimbert1974} by linear 
extrapolation of the 
Y$-$ O/H correlation to O/H=0, where Y and O/H, are respectively the
$^4$He mass fraction and the oxygen abundance relative to hydrogen
of a sample of dwarf irregular and BCD galaxies. Based on a relatively large
sample of 45 BCDs, Izotov \& Thuan (1998) found 
Y$_{p,IT}=0.2443\pm0.0015$ and a scaling $\di {\rm Y}/\di{\rm (O/H)}=45\pm19$. 

\smallskip
\noindent
The Pop III star comoving density as a function of redshift, $\rho_\star(z)$, 
is given by 
\q\label{eq:rhostar}
\rho_{\star}(z)=f_{\star}\frac{\Omega_b}{\Omega_M} 
\int^{\infty}_{M_{min}(z)}  n(M_{h},z) M_{h} \di M_{h},
\nq
\noindent
where $n(M_{h},z)$ is the comoving number density of dark matter halos of mass
$M_{h}$ at redshift $z$ given by the Press \& Schechter formalism \cite{Press1974}; 
the integral
gives the dark matter mass per unit volume contained in dark matter halos with
mass greater than $M_{min}(z)$, where $M_{min}$ (computed by 
Fuller \& Couchman (2000)) is a cutoff mass below which halos
cannot form stars due to the lack of cooling. $\Omega_M$ and $\Omega_b$ are the 
total matter and baryon density\footnote{ 
We adopt the `concordance' model values for the cosmological parameters: 
$h=0.7$, $\Omega_M=0.3$, $\Omega_{\Lambda}=0.7$, $\Omega_b=0.038$, 
$\sigma_8=0.9$, and $\Gamma=0.21$, where $h$ is the dimensionless Hubble 
constant, $H_0=100h$ km s$^{-1}$ Mpc$^{-1}$} 
in units of the critical density $\rho_c=3H_0^2/8\pi G$.
The ratio $\Omega_b/\Omega_M$ converts the 
integral into baryonic mass which is then turned into stars with an 
efficiency $f_{\star}$.
The latter is constrained by the contribution of the redshifted
light of Pop III stars to the NIRB. The observational data in the NIR (see e.g.
Hauser \& Dwek (2001) and references therein) are well fitted by a model in 
which the contribution from Pop III stars is added to the ``normal'' galaxy 
background light \cite{Salvaterra2002}, where the Pop III contribution is
given by 

\begin{figure}
\centerline{{
\psfig{figure=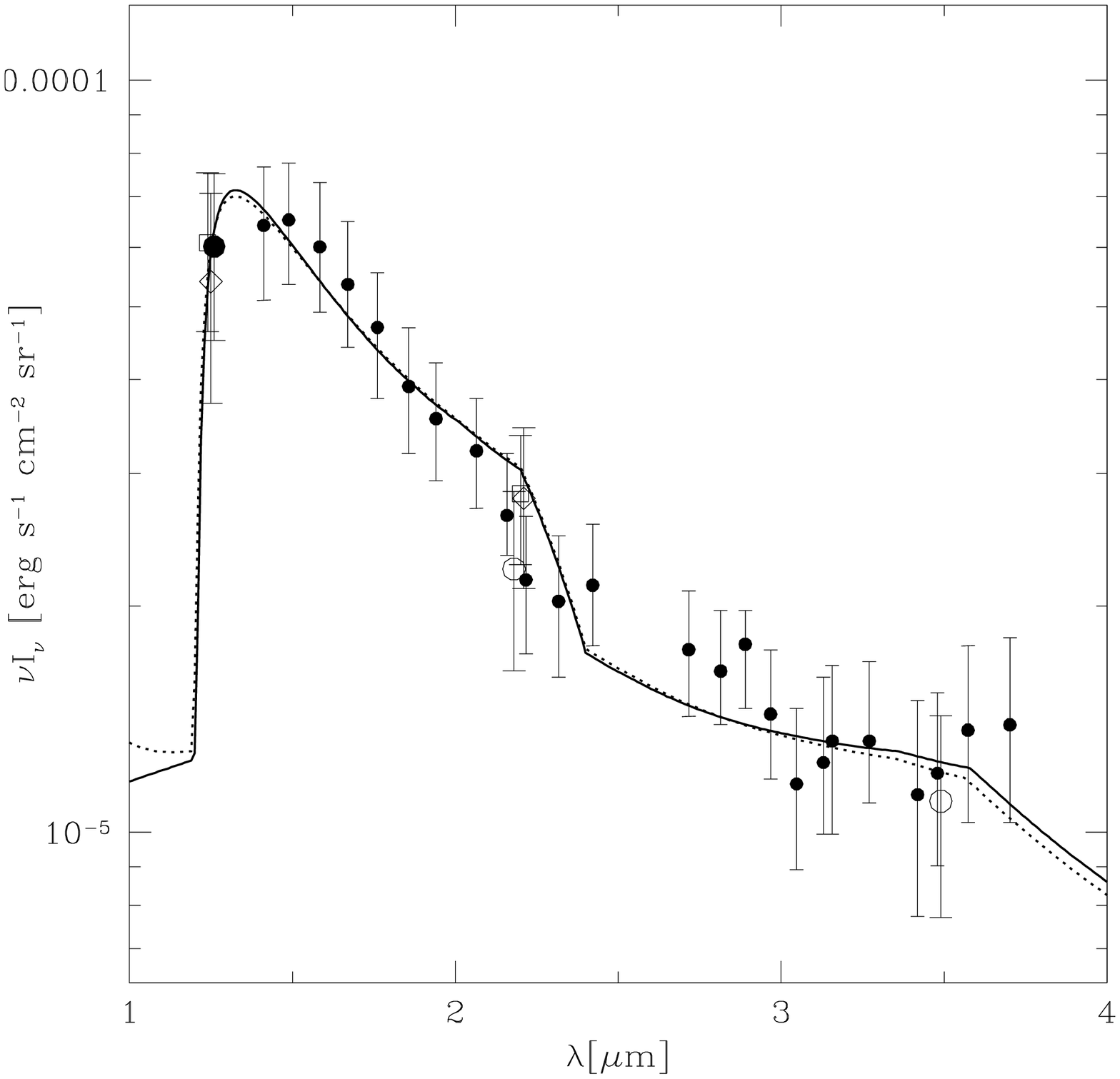,height=14cm}
}}
\caption{}
\label{fig:nirb}
\end{figure}

\q
I(\nu_{0},z_{0})= \frac{1}{4\pi}\int^{\infty}_{z_{0}} l_{\nu}(z,\phi) \rho_{\star}(z) e^{-\tau_{eff}(\nu_{0},z_{0},z)}\frac{dl}{dz}dz, 
\nq

\noindent
with  $l_\nu(z,\phi)$ being the specific luminosity of the population at
redshift $z$, computed using the spectra for metal-free stars obtained by
Schaerer (2002), $\phi$ is the initial mass function (IMF), 
$\nu=\nu_0(1+z)/(1+z_0)$, $\tau_{eff}(\nu_{0},z_{0},z)$ is the
effective optical depth at $\nu_0$ of the IGM between redshift $z_0$ and
$z$, and $dl/dz$ is the proper line element. 
The Pop III star formation efficiency required to  
account for the observed NIRB depends essentially only on IMF for those 
objects, the redshift $z_{end}$ at which 
the formation of Pop III stars ends being constrained to be $z_{end}\simeq 8.8$ by 
the J band data \cite{Salvaterra2002}. In addition to the NIRB, an early
star formation may contribute to the metal enrichment of the intergalactic 
medium (IGM). Although the enrichment history of the IGM at high redshift is
very uncertain, mainly depending on the metal escape fraction from galaxies 
and their actual spatial distribution, it sounds reasonable to adopt the mean
Ly$\alpha$ forest metallicity as a `reliable' first-order estimate of the
mean IGM metallicity.
Ly$\alpha$ forest QSO absorption line experiments indicate a mean IGM 
metallicity of $\log(Z_{IGM}/Z_\odot)=-2.5\pm0.5$ at $z\sim 3$
(Dav\'e et al. 1998, Telfer et al. 2002).  
Salvaterra \& Ferrara (2002) have shown that this  constraint
is  not exceeded by first stars if either {\bf [i]} their primordial Initial Mass
Function is a Salpeter one ($\phi(M)\propto M^{-2.35}$),
in which pair instability supernovae (SN$_{\gamma\gamma}$, i.e. progenitor stars with
mass in the range 130-260 $\Msun$ \cite{Heger2002}) are the dominant sources of heavy elements, 
or {\bf [ii]} only stars with masses in excess of 260 $M_\odot$ are
formed, that lock  their nucleosynthetic products into a very massive
black hole (VMBH) remnant \cite{Schneider2002}.
In the first case, the NIRB is well fitted with $f_\star=0.534$ and 
$z_{end}=8.788$. For the second case, although the model seems to overproduce
the VMBH density to respect of the estimate density of super massive black hole
(SMBH) (Merritt \&  Ferrarese 2001), at the moment there is no strong reason 
to reject this model, since the connection between VMBHs and SMBHs is not
clear. We adopted here $\phi(M)=\delta(1000\;\Msun)$
and the NIR data are reproduced with $f_\star=0.043$ and $z_{end}=8.830$; 
we will use these $z_{end}$ values in the following. 
Note that, as stars more massive than about $300 M_\odot$ have virtually the
same spectrum, the $\delta$-function distribution does not represent an 
unrealistic assumption \cite{Bromm2001}. 
These two best fit models to the NIRB data are shown in Figure~\ref{fig:nirb}.
Obviously, the NIRB fraction accounted for by PopIII stars is dircetly proportional
to $f_\star$; as a consequence the same linear dependence is found for $\Delta {\rm Y}_p$,
as it is clear from eq. 3 below and eq. 1.

\smallskip 
\noindent
The constraints on the IMF obtained from this fit can then be used to 
estimate the contribution of Pop III stars to $^4$He primordial abundance.
This is given by
\q\label{eq:yp}
\Delta {\rm Y}_p (\phi,z) = f_{He}(\phi)\;\Omega_{\star}(\phi,z)/\Omega_b
\nq
\noindent 
where $\Omega_{\star}(z)=\rho_{\star}/\rho_c$; 
$f_{He}$ is the $^4$He fraction of the stellar mass ejected via supernova explosions 
and/or through 
mass-loss in a radiatively-driven wind (although Kudritzki (2002), has 
hinted that winds might not be effective at $Z=0$, recent results
show that winds can take place even in metal-free stars;
Lentz, Hauschildt, Aufdenberg \& Baron, in preparation);
its value depends on the adopted IMF, $\phi$. 
The $^4$He produced by SN$_{\gamma\gamma}$ explosions of Pop III stars 
has  been computed by Heger \& Woosley (2002). 
Stars with mass above 260 $\Msun$ 
are likely to end up in VMBHs without $^4$He ejection.
On the other hand, according to very recent calculations \cite{Marigo2002}, wind mass-loss 
is significant 
only for very massive objects (750-1000 $\Msun$). For 
a 1000 $\Msun$ non rotating star, 17.33\% (0.03\%) of the initial mass 
is ejected via mass-loss as helium (oxygen) during the stellar lifetime. A
small amount of carbon and nitrogen is ejected as well. If 
rotation is taken into account the percentages are 16.97\% (He) and 0.15\%
(O). Pulsational instabilities may not have sufficient time to drive 
appreciable mass-loss in metal free stars \cite{Baraffe2001}.

\begin{figure}
\centerline{{
\psfig{figure=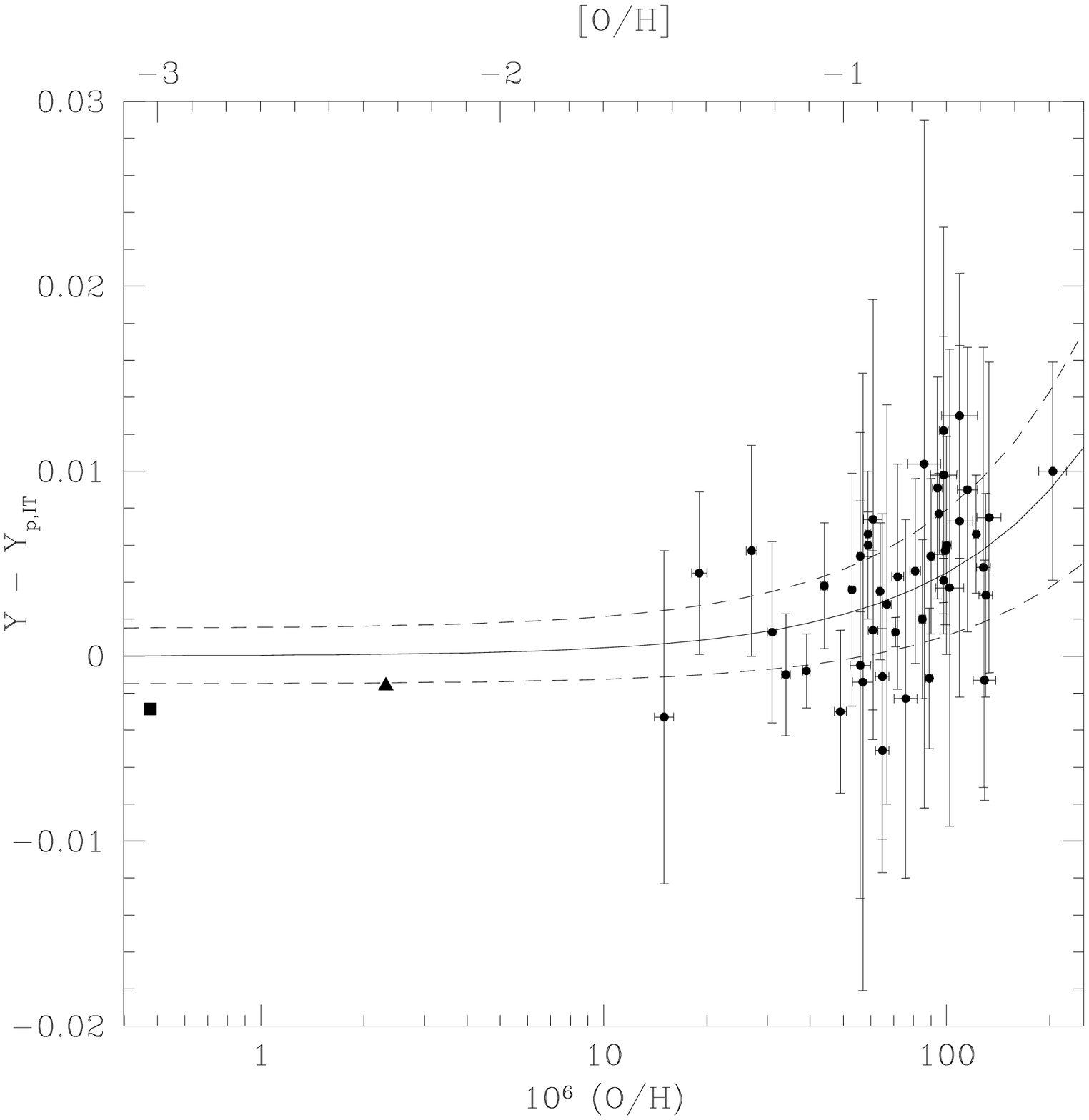,height=14cm}
}}
\caption{}
\label{fig:he}
\end{figure}

\noindent
We calculate the contribution of Pop III stars
to the $^4$He abundance according to eq.~(\ref{eq:yp}) down to redshift $z_{end}$ at which the formation of Pop III
stars ends. For the Salpeter IMF (case [i] above) $\Delta {\rm Y}_p$ is quite small, 
$\approx 2\times10^{-6}$ 
if mass-loss is not considered. This value doubles if one accounts for He
ejected from 750-1000 $\Msun$ stars via wind losses; inclusion of rotation
does not change these conclusions. 
In case [ii] there is no contribution from 
SN$_{\gamma\gamma}$, since 1000 $\Msun$ stars fall above the 
SN$_{\gamma\gamma}$ progenitor range; as already mentioned, mass-loss in a wind
is instead quite efficient at ejecting $^4$He. 
The resulting contribution is $\Delta {\rm Y}_p = 3.191\times10^{-3}$ 
(3.258$\times10^{-3}$) in the 
rotating (non-rotating) case. Moreover, the gas is also enriched in 
oxygen\footnote{Following common notation, we use square brackets to express
values in units of solar abundance, [O/H]=log(O/H) -- log(O/H)$_\odot$, and use
(O/H)$_\odot=5\times 10^{-4}$, where (O/H) is the abundance ratio of the two
species by number.} to 
[O/H]$=-2.33\; (-3.02)$, leading to a mean IGM metallicity of 
$\log(Z/Z_\odot)=-2.52\; (-3.15)$. In Figure~\ref{fig:he} we show the
$^4$He data points from the Izotov \& Thuan sample \cite{Izotov1998} 
from which we have subtracted the best fit primordial value, Y$_{p,IT}$. 
Our results for the non-rotating and rotating cases are also indicated
in the Figure by the square and triangle  points.  
Both points fall beyond the 1$\sigma$ error line (and the non-rotating one
beyond 2$\sigma$). Hence the conclusion is that a measure of the $^4$He
primordial abundance extrapolated to zero metallicity from the values
of currently sampled [O/H]$\simgt\;-1.6$ would include the 
contribution from the very first generation of stars and therefore 
overestimate the amount of `truly' primordial, BBN $^4$He. 
As a result, a correct determination of BBN $^4$He can only be made by sampling a 
strictly metal-free gas: any metal-enriched gas will already bear the 
signature of the $^4$He production of Pop III stars. 
A similar conclusion, although based on totally different arguments, has
been suggested in the past \cite{Mathews1993}.
We note that 
the exact horizontal position of the theoretical points is affected by 
the systematic uncertainty introduced by the implicit assumption made
of a homogeneous distribution of the stellar nucleosynthetic products;
such error is difficult to estimate. In general, as 
observations are pushed to lower metallicity gas, we expect the scattering
in the data to increase as a result of the inhomogeneous distribution of
oxygen, which is presumably enhanced close to the most actively star forming
regions of the universe. 
We reiterate that the value for $\Delta Y_p$ here inferred has to be considered as an 
upper limit, since a broader IMF in the range 300-1000 $\Msun$ leads to a 
lower contribution.

The inferred value of $\Delta$Y$_p$ gives us a measure of the accuracy with
which Y$_p$ can be measured though current observations 
of this type. In the most extreme case we can expect Y$_p = {\rm Y}_{p,IT} - \Delta
{\rm Y}_p = 0.2443 - 3.258\times 10^{-3}= 0.241$. Is this low value at odd with 
BBN theory~? It is well known  that there is a tension between
theoretical BBN predictions and observed abundances, in the sense that 
the latter typically fall short, particularly for $^4$He and $^7$Li. 
The further decrease of $^4$He abundance suggested here would aggravate 
the discrepancy, a fact which should 
lead us to consider even more seriously non-standard BBN models.  
Similar arguments apply if one considers D and $^4$He data simultaneously: 
observations of high-$z$ QSO absorbers seem to suggest a low primordial 
deuterium abundance, implying high nucleon-to-baryon ratio and 
Y$_p$ values. The contribution from Pop III stars will instead tend to decrease
Y$_p$, a problem which would be exacerbated if we are to believe 
lower (with respect to the IT one) determinations of such quantity, 
Y$_p= 0.234 \pm 0.003$, obtained by other groups \cite{OSS1997}. 
A lower Y$_p$ value would require a slower expansion of the radiation-dominated
universe. This behavior is predicted by a class of so-called ``extended quintessence''
models, as a result of a decrease of the effective gravitational constant $G$. 
Obviously, there is a lot that we can learn from BBN element abundance
measurements.
		   
\vskip 1.0truecm
\noindent
We thank E. Lentz, P. Marigo and their groups for providing results
in advance of publication; R. Schneider, E. Skillman and M. Tosi 
for useful comments and discussions.
This work was partially supported (AF) by the Research and Training Network
`The Physics of the Intergalactic Medium' set up by the European Community
under the contract HPRN-CT2000-00126 RG29185. RS acknowledges the Italian
MIUR for financial support.

\newpage

\newpage

\noindent
{\bf Figure 1 Caption}

\noindent
Best fit results of the NIRB data \cite{Salvaterra2002}. {\it Dotted line}: Salpeter IMF $z_{end}=8.788$, $f_\star=0.534$. {\it Solid line}: $\phi=\delta(1000\;\Msun)$ 
$z_{end}=8.83$ $f_\star=0.043$. 
The small filled circles are the NIRS data \cite{Matsumoto2000}. The open symbols
are the DIRBE results: squares for Wright (2001), 
diamonds for Cambr\'esy et al. (2001), and circles for 
Gorjian et al.(2000). 
The big filled circle is the Kiso star count measurement.
The errors are at 1$\sigma$ and for
all the data the Kelsall's model \cite{Kelsall1998} for the zodiacal light is applied.
The ``normal'' galaxy contribution is estimated from the {\it Subaru Deep Field}
 \cite{Totani2001}. For $\lambda>2.2$ $\mu$m a zero contribution from normal
galaxies to the NIRB is assumed.

\newpage

\noindent
{\bf Figure 2 Caption}

\noindent
Effect of Pop III stars on the $^4$He primordial abundance determination. 
The plot shows the 
$^4$He abundance, Y, to which the estimate by Izotov \& Thuan, 
Y$_{p,IT}$ \cite{Izotov1998} has
been subtracted, as a function of the oxygen content of the galaxy. 
The triangle (square) represents the $^4$He abundance obtained taking into 
account the helium ejected via mass-loss from  
rotating (non-rotating) Pop III stars. The black dots are
the data of the sample of Izotov \& Thuan (1998). The solid line is 
the best fit relation Y -- O/H and the dashed lines represent the 
corresponding 1$\sigma$ errors \cite{Izotov1998}.
\newpage

\end{document}